Optimized Tersoff and Brenner empirical potential parameters for lattice dynamics and phonon thermal transport in carbon nanotubes and graphene

L. Lindsay

Department of Physics, Boston College, Chestnut Hill, MA 02467, USA

Department of Physics, Computer Science, and Engineering, Christopher Newport University, Newport News, VA 23606, USA

D. A. Broido

Department of Physics, Boston College, Chestnut Hill, MA 02467, USA

**Abstract**

We have examined the commonly used Tersoff and Brenner empirical interatomic potentials in the context of the phonon dispersions in graphene. We have found a parameter set for each empirical potential that provides improved fits to some structural data and to the in-plane phonon dispersion data for graphite. These optimized parameter sets yield values of the acoustic phonon velocities that are in better agreement with measured data. They also provide lattice thermal conductivity values in single-walled carbon nanotubes that are considerably improved compared to those obtained from the original parameter sets.

I **INTRODUCTION**

Numerous theoretical investigations of the thermal transport properties of carbon based materials such as diamond [1-3], graphite [4-5], graphene [6-9] and carbon nanotubes [10-21] have been performed in recent years partly because these materials possess the highest known thermal conductivities [22-30]. Theoretical investigations of phonon thermal transport in materials commonly employ either molecular dynamics (MD) simulations or Boltzmann transport approaches. In both cases accurate representation of the interactions between atoms is required. Empirical interatomic potentials (EIPs) are frequently employed because they provide these interactions in a conveniently extractable form. The most commonly used EIPs in single-walled carbon nanotubes (SWCNTs) and graphene are those developed by Tersoff [31, 32] and Brenner [33, 34]. The lattice thermal conductivity depends strongly on the phonon dispersions and near-zone-center velocities in these systems. This is especially true in carbon based crystals such as diamond, graphite, graphene and SWCNTs because their extremely high Debye temperatures. Thus, an accurate description of the lattice dynamics is critically important in modeling the lattice thermal conductivities of SWCNTs and graphene.

The original parameter sets of the Tersoff and Brenner EIPs do not accurately reproduce the phonon dispersions of graphene, as has been noted previously [35]. In particular, they do not accurately obtain the velocities of the three acoustic branches near the center of the Brillouin zone, thus, also misrepresenting these properties in SWCNTs. We present here optimized parameter sets for the Tersoff and Brenner EIPs, which better represent the lattice dynamical properties of graphene. We test these optimized parameter sets on the lattice thermal conductivites of SWCNTs and find them to yield values that are in much better agreement with available data. In section II we briefly describe the Tersoff and Brenner EIPs and the approach we took in optimizing the empirical parameters. Section III presents the optimized parameter sets and compares the graphene phonon dispersions obtained from the original and optimized parameter sets. It also examines the phonon thermal conductivities of SWCNTs obtained from old and new parameter sets for both EIPs. Section IV presents a summary and conclusions.

**II THEORY**

The Tersoff and Brenner EIPs are convenient, short-range, bond-order, empirical potentials which are often used in MD simulations and other calculations to model different properties of carbon-based materials [2, 9-10, 15-21]. The convenience of these potentials comes from their rather simple, analytical forms and the short-range of atomic interactions. For carbon-based systems, the Tersoff model has nine adjustable parameters (listed in Table 1) that were originally fit to cohesive energies of various carbon systems, the lattice constant of diamond, and the bulk modulus of diamond. The Brenner EIP is based directly on the Tersoff EIP but has additional terms and parameters which allow it to better describe various chemical reactions in hydrocarbons and include nonlocal effects (parameters listed in Table 2).

*Tersoff model*

The analytical form for the pair-potential, $V_{ij}$, of the Tersoff model is given by the following functions with corresponding parameters listed in Table 1 [31, 32]:

$$V_{ij} = f_{ij}^C \left( a_{ij} f_{ij}^R - b_{ij} f_{ij}^A \right) \tag{1a}$$

$$f_{ij}^R = A e^{-\lambda_1 r_{ij}} \tag{1b}$$

$$f_{ij}^A = Be^{-\lambda_2 r_{ij}} \tag{1c}$$

where $r_{ij}$ is the distance between atoms $i$ and $j$, $f_{ij}^A$ and $f_{ij}^R$ are competing attractive and repulsive pair-wise terms and $f_{ij}^C$ is a cut-off term which ensures only nearest neighbor interactions. The term, $a_{ij}$, is a range-limiting term on the repulsive potential that is typically set equal to one. We do so here. The bond angle term, $b_{ij}$, depends on the local coordination of atoms around atom $i$ and the angle between atoms $i$, $j$, and $k$:

$$b_{ij} = \left(1 + \beta^n \zeta_{ij}^n\right)^{-1/2n} \tag{2a}$$

$$\zeta_{ij} = \sum_{k \neq i,j} f_{ik}^C g_{ijk} e^{\lambda_3^3 (r_{ij} - r_{ik})^3} \tag{2b}$$

$$g_{ijk} = 1 + \frac{c^2}{d^2} - \frac{c^2}{d^2 + (h - \cos[\theta_{ijk}])^2} \tag{2c}$$

where $\theta_{ijk}$ is the angle between atoms $i$, $j$, $k$. This bond angle term allows the Tersoff model to describe the strong covalent bonding that occurs in carbon systems, which cannot be represented by purely central potentials. This angle-dependent term also allows for description of carbon systems that bond in different geometries, such as tetrahedrally-bonded diamond and 120° tri-bonded graphene.

$A = 1393.6\ eV$         $B = 346.74\ eV$
$\lambda_1 = 3.4879\ Å^{-1}$     $\lambda_2 = 2.2119\ Å^{-1}$
$\lambda_3 = 0.0000\ Å^{-1}$     $n = 0.72751$
$c = 38049.0$            $\beta = 1.5724 \times 10^{-7}$
$d = 4.3484$             $h = -0.57058$
$R = 1.95\ Å$            $D = 0.15\ Å$

Table 1   Original parameters for the Tersoff EIP for carbon-based systems given in Ref. 32.

*Brenner model*

The Brenner potential for solid-state carbon structures is given by the following functions with corresponding parameters listed in Table 2 [33, 34]:

$$V_{ij} = f_{ij}^C \left( f_{ij}^R - \bar{b}_{ij} f_{ij}^A \right) \tag{3a}$$

$$f_{ij}^R = \left(1 + \frac{Q}{r_{ij}}\right) A e^{-\alpha r_{ij}} \tag{3b}$$

$$f_{ij}^A = \sum_{n=1}^{3} B_n e^{-\lambda_n r_{ij}} \tag{3c}$$

where many of the terms are similar to the Tersoff model described above and the bond angle term, $\bar{b}_{ij}$, is given by:

$$\bar{b}_{ij} = \frac{1}{2}\left(b_{ij}^{\sigma-\pi} + b_{ji}^{\sigma-\pi}\right) + \Pi_{ij}^{RC} + b_{ij}^{DH} \tag{4a}$$

$$b_{ij}^{\sigma-\pi} = \left(1 + \sum_{k \neq i,j} f_{ik}^C g_{ijk}\right)^{-1/2} \tag{4b}$$

$$g_{ijk} = \sum_{i=0}^{5} \beta_i \cos^i[\theta_{ijk}] \tag{4c}$$

Here, $b_{ij}^{\sigma-\pi}$ depends on the local coordination of atoms around atom *i* and the angle between atoms *i*, *j*, and *k*, $\theta_{ijk}$. This pi-bond function is symmetric for graphene, graphite, and diamond, $b_{ij}^{\sigma-\pi} = b_{ji}^{\sigma-\pi}$. The coefficients, $\beta_i$, in the bond-bending spline function, $g_{ijk}$, were fit to experimental data for graphite and diamond and are also listed in Table 2. The term, $\Pi_{ij}^{RC}$, accounts for various radical energetics, such as vacancies, which are not considered here; thus, this term is taken to be zero. The term, $b_{ij}^{DH}$, is a dihedral bending function that depends on the local conjugation and is zero for diamond but important for describing graphene and SWCNTs. This dihedral function involves third nearest-neighbor atoms and is given by [33, 34, 36]:

$$b_{ij}^{DH} = \frac{T_0}{2} \sum_{k,l \neq i,j} f_{ik}^C f_{jl}^C \left(1 - \cos^2[\Theta_{ijkl}]\right) \qquad (5)$$

where $T_0$ is a parameter, $f_{ij}^C$ is the cut-off function, and $\Theta_{ijkl}$ is the dihedral angle of four atoms identified by the indices, $i,j,k,$ and $l$, and is given by:

$$\cos[\Theta_{ijkl}] = \vec{\eta}_{jik} \cdot \vec{\eta}_{ijl} \qquad (6a)$$

$$\vec{\eta}_{jik} = \frac{\vec{r}_{ji} \times \vec{r}_{ik}}{|\vec{r}_{ji}||\vec{r}_{ik}|\sin[\theta_{ijk}]} \qquad (6b)$$

where $\vec{\eta}_{jik}$ and $\vec{\eta}_{ijl}$ are unit vectors normal to the triangles formed by the atoms given by the subscripts, $\vec{r}_{ij}$ is the vector from atom $i$ to atom $j$, and $\theta_{ijk}$ is the angle between atoms $i, j,$ and $k$. In flat graphene, the dihedral angle, $\Theta_{ijkl}$, is either 0 or $\pi$ and the dihedral term is subsequently zero [36]. Bending of the graphene layer leads to a contribution from this term.

$$A = 10953.544162170 \, eV \qquad B_1 = 12388.79197798 \, eV$$
$$B_2 = 17.56740646509 \, eV \qquad B_3 = 30.71493208065 \, eV$$
$$\alpha = 4.7465390606595 \, \text{Å}^{-1} \qquad \lambda_1 = 4.7204523127 \, \text{Å}^{-1}$$
$$\lambda_2 = 1.4332132499 \, \text{Å}^{-1} \qquad \lambda_3 = 1.3826912506 \, \text{Å}^{-1}$$
$$Q = 0.3134602960833 \, \text{Å} \qquad R = 2.0 \, \text{Å}$$
$$D = 1.7 \, \text{Å} \qquad T_0 = -0.00809675$$

$$\beta_0 = 0.7073 \qquad \beta_1 = 5.6774$$
$$\beta_2 = 24.0970 \qquad \beta_3 = 57.5918$$
$$\beta_4 = 71.8829 \qquad \beta_5 = 36.2789$$

Table 2  Original parameters for the Brenner EIP for solid state carbon-based structures given in Ref. 34. Also listed are the coefficients for the fifth-order polynomial spline, $g_{ijk}$, described in Ref. 34.

Some of the main differences when compared to the Tersoff EIP are: the Brenner EIP includes two additional exponential terms with corresponding adjustable parameters in the attractive pair-wise term, it includes a screened Coulomb term in the repulsive pair-wise term, it uses a fifth-order polynomial spline between bond orders for diamond and graphite, and it includes a dihedral bending term for bond energies which plays a role in SWCNTs and graphene.

*Parameter optimization*

We have implemented a $\chi^2$ minimization procedure for each of these EIPs. A numerical algorithm was developed to minimize $\chi^2$ given by [37, 38]:

$$\chi^2 = \sum_i \frac{(\eta_i - \eta_{\exp})^2}{\eta_{\exp}^2} \zeta_i \qquad (7)$$

where $\eta_{\exp}$ are experimental parameters used in the fitting process, $\eta_i$ are the corresponding values obtained using each potential, and $\zeta_i$ are weighting factors that determine the relative importance of $\eta_i$ in the fitting procedure.

In minimizing $\chi^2$, the greatest significance was given to the phonon frequencies, $\omega_\lambda$, and the zone-center acoustic velocities, $v_\lambda = \partial \omega_\lambda / d\vec{q}$, of graphene in the high-symmetry directions. Here, $\lambda = (\vec{q}, j)$ designates a phonon with wavevector, $\vec{q}$, in branch, $j$. The phonon frequencies are determined by diagonalization of the dynamical matrix for a given $\vec{q}$ in the two-dimensional graphene Brillouin zone. The dynamical matrix is:

$$D_{\alpha\beta}^{\kappa\kappa'}(\vec{q}) = \frac{1}{\sqrt{m_\kappa m_{\kappa'}}} \sum_{\ell'} \Phi_{\alpha\beta}^{0\kappa,\ell'\kappa'} e^{i\vec{q}\cdot\vec{R}_{\ell'}} \qquad (8)$$

where $\ell\kappa$ designates the $\kappa^{th}$ atom in the $\ell^{th}$ unit cell, $m_\kappa$ is the mass of the $\kappa^{th}$ atom, $\vec{R}_\ell$ is the lattice vector for the $\ell^{th}$ unit cell, and $\alpha$ and $\beta$ are Cartesian components. In Eq. 8, $\Phi_{\alpha\beta}^{0\kappa,\ell'\kappa'}$ are second-order interatomic force constants (IFCs) which are determined by each EIP.

We attached the most significance to the phonon frequencies and zone-center acoustic velocities because of the important roles that they play in thermal transport calculations. The parameters calculated by each EIP for graphene were compared to the corresponding experimental parameters for in-plane graphite. First principles calculations of the phonon dispersions in graphene have found excellent agreement with measured in-plane dispersion of graphite [39]. This is consistent with the known weak coupling between the graphene layers in graphite. The cohesive energies and lattice constants of graphite and diamond were also considered in the minimization procedure, but were given lesser weight.

Using this minimization procedure for the Tersoff EIP, we found that simply modifying the *h* parameter, which helps to adjust the strength of the bond-angle term, provided vast improvement to the optical branches of the phonon dispersion, while also improving the fit to the transverse acoustic (*TA*) zone-center velocity. Adjustment of the *B* parameter associated with the strength of the attractive term was required to retain a decent fit to the experimental lattice constants for both graphite and diamond. Simple adjustment of the *h* parameter has previously been found to significantly improve the fits to the measured linear expansion coefficient [40] and thermal conductivity of bulk silicon [41].

A slightly different approach was taken in optimizing the Brenner model to better fit the phonon spectrum. The Brenner potential includes a dihedral bonding term, not included in the Tersoff model, which plays a role in graphene systems. The dihedral term, Eqn. 5, has a single adjustable parameter, $T_0$, which was determined by fitting the lattice constant of a hypothetical three-dimensional, hexagonal system whose dihedral bond angles are $\pi/2$ [34, 42]. Changing this parameter alone simply alters the out-of-plane acoustic and optic (*ZA* and *ZO*) modes in graphite, but leaves the other phonon modes unaltered and has no effect on diamond since $T_0$ is zero for the tetrahedral configuration. We have chosen to adjust this parameter to better fit the phonon frequencies for the *ZA* branch in graphite.

The six coefficients, $\beta_i$, in the fifth-order polynomial spline, $g_{ijk}$, (Eq. 4c) used to represent the bond-bending term in the Brenner EIP introduce additional flexibility not available in the Tersoff potential. These coefficients were originally determined by fixing the values of $g_{ijk}$ and their first and second derivatives at 109° and 120° (corresponding to diamond and graphite bond angles) to match various experimental data [34]. The values for $g_{ijk}$ were determined by fitting the cohesive bond energies of graphite and diamond while the second derivatives were chosen to fit the elastic constants, $c_{11}$, for diamond and in-plane graphite. The first derivatives were simply chosen to suppress oscillations of the spline function. The coefficients, $\beta_i$, are determined from these values for $g_{ijk}$. These coefficients fix the spline and its derivatives at the bond angles for diamond and graphite, and the function is interpolated between these angles. Thus, they can be adjusted to separately fit experimental data for graphite, while leaving the representation for diamond unaltered. The bond angles for SWCNTs, though close to graphite, fall between these two structures and thus depend on both. We note that the bond angles for large diameter SWCNTs approach that of graphite.

We restricted the parameter optimization of the Brenner potential to $T_0$ and $\beta_i$ to avoid significantly altering the previous fits to the extensive structural experimental data sets [33,34]. The values for $g_{ijk}$ at 109° and 120° were altered slightly to better fit the given experimental lattice constants for diamond and graphite. We then adjusted the values of the second derivatives of $g_{ijk}$ in these materials to better fit the zone-center acoustic velocities and corresponding phonon frequencies. Upon plotting the fifth order polynomial spline given by the old coefficients versus the optimized coefficients no visual difference can be seen. However, these small changes introduce noticeable changes in the corresponding phonon dispersions in graphene.

*Tersoff*

$$h = -0.930 \qquad B = 430.0\, eV$$

*Brenner*

$$T_0 = -0.0165$$
$$\beta_0 = 0.0000 \qquad \beta_1 = -3.1822$$
$$\beta_2 = -19.9928 \qquad \beta_3 = -51.4108$$
$$\beta_4 = -61.9925 \qquad \beta_5 = -29.0523$$

Table 3  Optimized parameters and coefficients for the Tersoff and Brenner EIPs. All parameters not listed are unaltered from the original sets.

**III RESULTS AND DISCUSSION**

The optimized parameter sets for the Tersoff and Brenner EIPs are listed in Table 3. These parameters provide improved fits to experimental phonon acoustic velocities and frequencies without significantly altering fits to other structural data. The calculated phonon dispersion for graphene as given by the Tersoff (Brenner) EIP is shown in Figure 1 (Figure 2) along with the corresponding measured in-plane phonon dispersion for graphite [43, 44]. In each figure the black (red) lines correspond to the optimized (original) parameter sets.

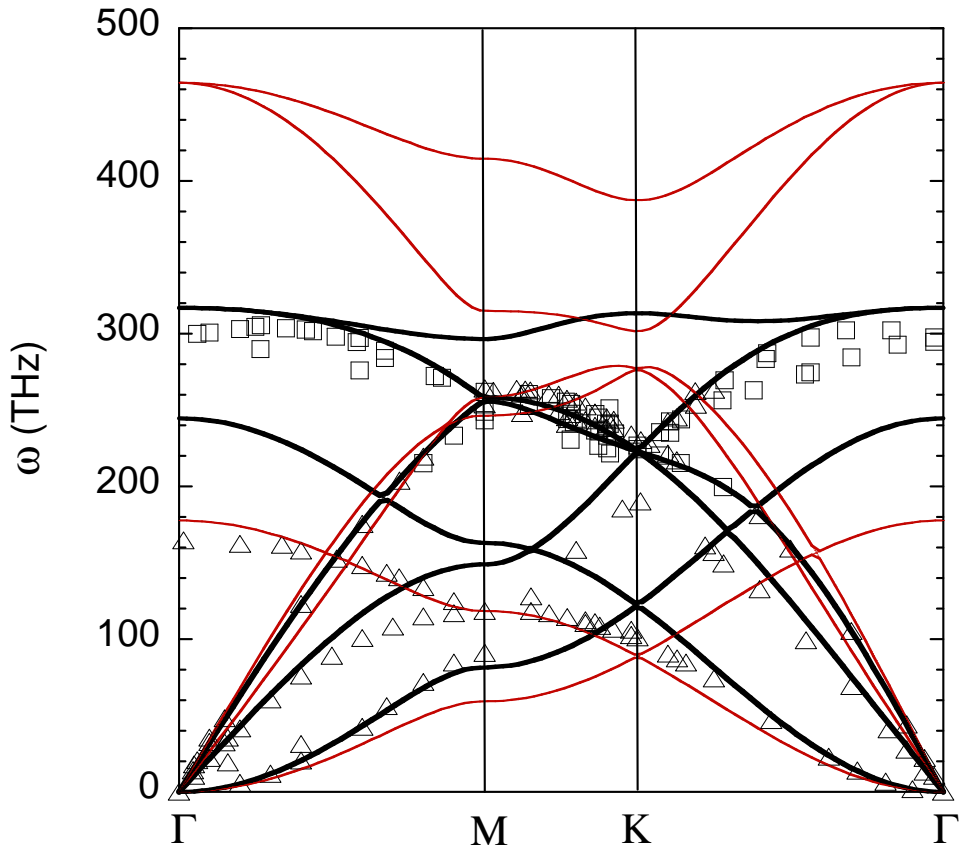

Figure 1   Phonon dispersion for graphene along high symmetry directions obtained using the Tersoff EIP. Thick black lines correspond to the optimized parameter set (this work); thin red lines correspond to the original parameter set. Squares (triangles) are in-plane experimental data points for graphite from Ref. 43 (Ref. 44).

The original set of parameters for the Tersoff EIP gives higher values for the *TA* branch velocities in the high-symmetry directions compared to the measured data, while giving values for the quadratic *ZA* frequencies that fall below experiment around the M point. The most obvious failure of the Tersoff EIP in describing the phonon dispersion of graphene comes in the highest optical branches, as seen in Figure 1. The measured in-plane upper optic modes at the $\Gamma$-point are degenerate with a value of 300 THz, while the Tersoff potential gives a value of 470 THz, a discrepancy of nearly 40%. The Tersoff model using the optimized parameter set more accurately describes these upper optic phonon branches while providing a decent fit to the acoustic velocities and phonon frequencies. However, using these parameters provides a poorer fit to the experimental dispersion for the out-of-plane *ZO* branch. The inability to simultaneously fit the acoustic branches and all of the optic branches may be a consequence of the Tersoff potential's short range, with only second nearest neighbor interactions represented. Tewary and Yang obtained a better fit to the phonon dispersion in graphene using a longer-range EIP that extended to fourth nearest neighbors [35]. Los *et al.* have also developed a somewhat more complicated long-range EIP based on the Brenner EIP, though, to the best of our knowledge, phonon dispersion results have not yet been published [45].

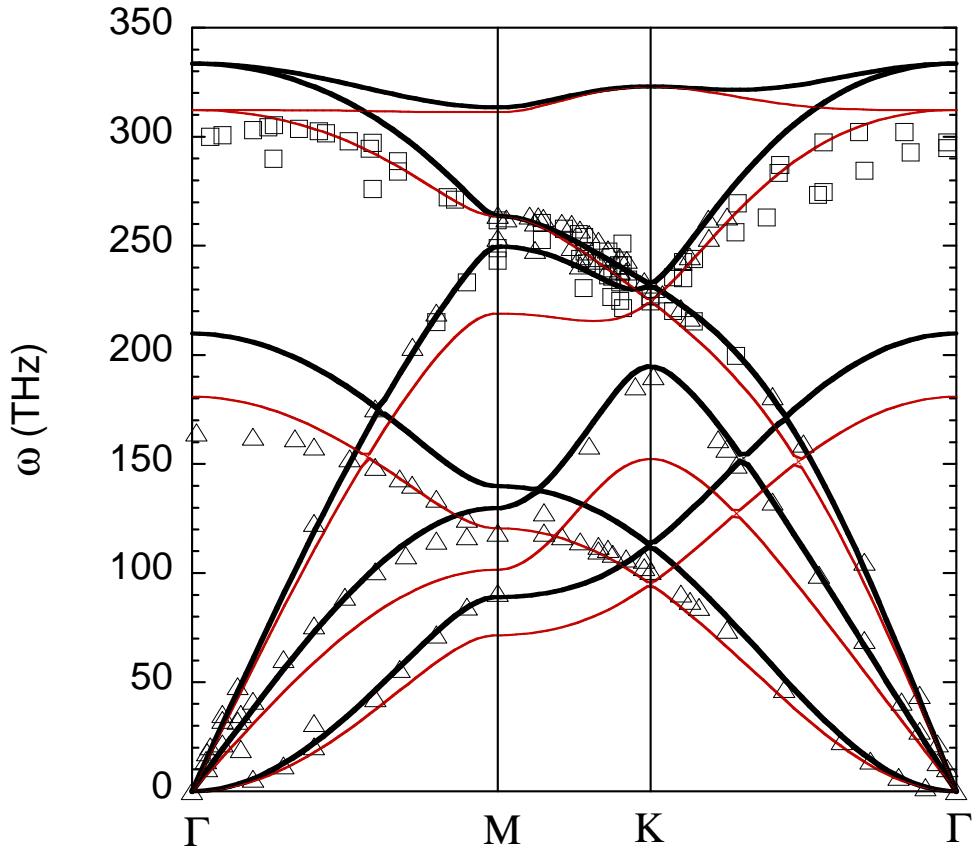

Figure 2  Phonon dispersion for graphene as given by the Brenner EIP for high symmetry directions. Thick black lines correspond to the optimized parameter set (this work); thin red lines correspond to the original parameter set. Squares (triangles) are in-plane experimental data points for graphite from Ref. 43 (Ref. 44).

The original set of Brenner parameters, while providing a much better description of the optic branches, does not accurately represent the zone-center velocities for all of the acoustic modes. With the original Brenner EIP parameters, the velocities of the *TA* branch in the high-symmetry directions are too low by 30%, those of the longitudinal acoustic (*LA*) branch 12% too small, and the dispersion of the *ZA* branch undershoots the data by as much as 20%. The optimized parameter set improves the fit to the zone-center acoustic velocities and most of the experimental *ZA*, *TA* and *LA* dispersion data. This optimized set provides a somewhat worse fit to the optic dispersion. The inability to simultaneously fit both acoustic and optic branches is again most likely a consequence of the short range of this EIP.

Table 4 lists measured lattice constants, cohesive energies and acoustic velocities in the $\Gamma \to M$ direction for graphite and the $\Gamma \to X$ direction for diamond compared with those obtained from the Tersoff and Brenner EIPs using both the original and optimized parameter sets.

|  | *Experiment* | *Tersoff original* | *Tersoff optimized* | *Brenner original* | *Brenner optimized* |
|---|---|---|---|---|---|
| $a_{lat}$ (Å) (gra) | 2.459[a] | 2.530 | 2.492 | 2.460 | 2.460 |
| $v_{TA}$ (m/s) (gra) | 14920[b] | 18274 | 14926 | 10641 | 12968 |
| $v_{LA}$ (m/s) (gra) | 21819[b] | 24002 | 21833 | 19388 | 20763 |
| $E_{coh}$ (eV) (gra) | -7.374[c] | -7.396 | -7.978 | -7.395 | -7.401 |
| $a_{lat}$ (Å) (dia) | 3.567[d] | 3.566 | 3.645 | 3.566 | 3.567 |
| $E_{coh}$ (eV) (dia) | -7.349[c] | -7.371 | -6.537 | -7.370 | -7.361 |

[a]Reference 46.
[b]From dispersion Reference 44.
[c]Reference 47.
[d]Reference 48.

Table 4  Lattice constant, cohesive energy, and acoustic phonon velocities for in-plane graphite and diamond as given by the Tersoff and Brenner EIPs using original and optimized parameter sets as compared to experiment.

For both the Tersoff and Brenner EIPs the optimized parameter sets provide better agreement with acoustic phonon velocities in graphite since these were included as fitting parameters. For the Tersoff EIP the optimized parameter set gives better agreement to the measured in-plane lattice constant of graphite but poorer agreement to the cohesive energy. For diamond, both the lattice constant and the cohesive energy are fit better by the original parameter set. For the Brenner EIP the optimized parameters leave the structural data for graphite and diamond unaltered due to the fitting procedure described above.

Recent calculations of the lattice thermal conductivity, $\kappa_L$, of SWCNTs based on a Boltzmann transport approach [14] have been performed with second- and third-order IFCs obtained from the Tersoff EIP with the original parameter set. Another recent molecular dynamics simulation of $\kappa_L$ [15] employed the Brenner EIP with original parameters. Each calculation using these different transport models obtained values for $\kappa_L$ that are inconsistently low compared to measured values [49, 50]. For example, for SWCNTs with lengths around 3μm and diameters in the 1-2nm range, the measured room temperature $\kappa_L$s were around 3000-3500W/m-K. In contrast, calculated room temperature values for (10,10) SWCNTs were in both cases [14, 15] considerably below 1000W/m-K. Employing the BTE approach developed previously by us [14], we have calculated the $\kappa_L$ for a 3μm (10,10) SWCNT at T=300K for each EIP using the original and optimized parameter sets. These values are listed in Table 5. It is evident that for both Tersoff and Brenner EIPs, the optimized parameter sets give values of $\kappa_L$ that are significantly larger and more consistent with those obtained experimentally. Part of the reason for this is that both potentials are too anharmonic as reflected in the third-order IFCs, so the phonon-phonon scattering rates are too large. It is interesting that the original Brenner parameters give extremely low $\kappa_L$, and in particular much lower than even the original Tersoff parameters. This occurs because the corresponding TA and LA acoustic velocities are much too low, as seen in Table 4, and because $\kappa_L$ depends on both phonon velocities and energies. To illustrate this sensitivity, we note that if the second-order IFCs from the optimized Brenner EIP are used with the third-order IFCs from the original parameter set, the resulting $\kappa_L$ jumps from 250W/m-K to 1080W/m-K.

|  | *original* | *optimized* |
|---|---|---|
| Tersoff $\kappa_L$ (W/m-K) | 600 | 1950 |
| Brenner $\kappa_L$ (W/m-K) | 250 | 2000 |

Table 5  Thermal conductivity of a 3μm (10,10) SWCNT at T=300K using the original and optimized parameter sets for the Tersoff and Brenner EIPs.

## IV SUMMARY AND CONCLUSIONS

Optimized parameter sets for the Tersoff and Brenner empircal interatomic potentials have been presented that provide overall improved agreement with the *ZA*, *TA* and *LA* phonon branches in graphene and in-plane graphite. In particular, the near-zone-center velocities of these branches are better fit by the optimized parameter sets. These optimized parameters for the Tersoff and Brenner potentials have been demonstrated to improve the agreement between the calculated lattice thermal conductivity of SWCNTs and measured values. Based on this, we expect that they will also provide better representations of the lattice dynamics of and modeling of phonon thermal transport in graphene, graphene nanoribbons, and graphite.

**ACKNOWLEDGEMENTS**

The authors gratefully acknowledge the National Science Foundation under Grant No. CBET 0651381, and the Donors of the American Chemical Society Petroleum Research Fund for support of this research. We also wish to thank Dr. Natalio Mingo for his helpful input on this work.

**REFERENCES**

[1]  J. Che, T. Çağin, W. Deng, W. Goddard III, J. Chem. Phys. **113**, 6888 (2000).

[2]  A. Sparavigna, Phys. Rev. B **65**, 064305 (2002).

[3]  A. Ward, D. A. Broido, D. A. Stewart, G. Deinzer, Phys. Rev. B **80**, 125203 (2009).

[4]  P. G. Klemens and D. F. Pedraza, Carbon **32**, 735 (1994).

[5]  P. G. Klemens, J. Wide Bandgap Mater. **7**, 332 (2000).

[6]  D. L. Nika, E. P. Pokatilov, A. S. Askerov, A. A. Balandin, Phys. Rev. B **79**, 155413 (2009).

[7]  D. L. Nika, S. Ghosh, E. P. Pokatilov, A. A. Balandin, Appl. Phys. Lett. **94**, 203013 (2009).

[8]  J. H. Seol, I. Jo, A. L. Moore, L. Lindsay, Z. H. Aitken, M. T. Pettes, X. Li, Z. Yao, R. Huang, D. A. Broido, N. Mingo, R. S. Ruoff, L. Shi, accepted to Science (2010).

[9]  J. Hu, X. Ruan, Y. P. Chen, Nano Lett. **9**, 2730 (2009).

[10]  J. X. Cao, X. H. Yan, Y. Xiao, and J. W. Ding, Phys. Rev. B **69**, 073407 (2004).

[11]  N. Mingo and D. A. Broido, Nano Lett. **5**, 1221 (2005).

[12]  J. Wang and J.-S. Wang, Appl. Phys. Lett. **88**, 111909 (2006).

[13]  Y. Gu and Y. Chen, Phys. Rev. B **76**, 134110 (2007).

[14]  L. Lindsay, D. A. Broido, N. Mingo, Phys. Rev. B **80**, 125407 (2009).

[15]  J. A. Thomas, R. U. Iutzi, A. J. H. McGaughey, Phys. Rev. B **81**, 045413 (2010).

[16]  S. Berber, Y.-K. Kwon, D. Tománek, Phys. Rev. Lett. **84**, 4613 (2000).

[17]  J. Che, T. Çağin, W. Goddard III, Nanotechnology **11**, 65 (2000).

[18]  C. W. Padgett and D. W. Brenner, Nano Lett. **4**, 1051 (2004).

[19]  Z. Yao, J.-S. Wang, B. Li, G.-R. Liu, Phys. Rev. B **71**, 085417 (2005).

[20]  D. Donadio and G. Galli, Phys. Rev. Lett. **99**, 255502 (2007).

[21]  J Lukes and H Zhong, J. Heat Transfer **129**, 705 (2007).

[22]  D. G. Onn, A. Witek, Y. Z. Qiu, T. R. Anthony, W. F. Banholzer, Phys. Rev. Lett. **68**, 2806 (1992).

[23]  J. R. Olsen, R. O. Pohl, J. W. Vandersande, A. Zoltan, T. R. Anthony, W. F. Banholzer, Phys. Rev. B **47**, 14850 (1993).

[24]  L. Wei, P. K. Kuo, R. L. Thomas, T. R. Anthony, W. F. Banholzer, Phys. Rev. Lett. **70**, 3764 (1993).

[25]  G. A. Slack, Phys. Rev. **127**, 694 (1962).

[26]  C. A. Klein and M. G. Holland, Phys. Rev. **136**, 575A (1964).

[27]  A. A. Balandin, S. Ghosh, W. Bao, I. Calizo, D. Teweldebrhan, F. Miao, C. Ning Lau, Nano Lett. **8**, 902 (2008).

[28]  S. Ghosh, I. Calizo, D. Teweldebrhan, E. P. Pokatilov, D. L. Nika, A. A. Balandin, W. Bao, F. Miao, and C. N. Lau, Appl. Phys. Lett. 92, 151911 (2008).

[29]  C. Yu, L. Shi, Z. Yao, D. Li, A. Majumdar, Nano Lett. **5**, 1842 (2005).

[30]  E. Pop, D. Mann, Q. Wang, K. Goodson, H. Dai, Nano Lett. **6**, 96 (2006).

[31]  J. Tersoff, Phys. Rev. Lett. **61**, 2879 (1988).

[32] J. Tersoff, Phys. Rev. B **37**, 6991 (1988).

[33] D. W. Brenner, Phys. Rev. B **42**, 9458 (1990).

[34] D. W. Brenner, O. A. Shenderova, J. A. Harrison, S. J. Stuart, B. Mi, S. B. Sinnott, J. Phys: Cond. Matt. **14**, 783 (2002).

[35] V. K. Tewary and B. Yang, Phys. Rev. B **79**, 075442 (2009).

[36] Q. Lu, M. Arroyo, R. Huang, J. Phys. D: Appl. Phys. **42**, 102002 (2009).

[37] W. H. Press, S. A. Teukolsky, W. T. Vetterling, B. P. Flannery, *Numerical Recipes in Fortran* (Cambridge, Cambridge University Press, 1992).

[38] L. Lindsay and D. A. Broido, J Phys.: Cond. Mat. **20**, 165209 (2008).

[39] N. Mounet and N. Marzari, Phys. Rev. B **71**, 205214 (2005).

[40] L. J. Porter, M. Yamaguchi, H. Kaburaki, M. Tang, J. App. Phys. **81**, 96 (1997).

[41] D. A. Broido, A. Ward, N. Mingo, Phys. Rev. B **72**, 014308 (2005).

[42] A. Y. Liu, M. L. Cohen, K. C. Hass, M. A. Tamor, Phys. Rev. B **43**, 6742 (1991).

[43] J. Maultzsch, S. Reich, C. Thomsen, H. Requardt, P. Ordejón, Phys. Rev. Lett. **92**, 075501 (2004).

[44] M. Mohr, J. Maultzsch, E. Dobardžić, S. Reich, I. Milošević, M. Damnjanović, A. Bosak, M. Krisch, C. Thomsen, Phys. Rev. B **76**, 035439 (2007).

[45] J. H. Los, L. M. Ghiringhelli, E. J. Meijer, A. Fasolino, Phys. Rev. B **72**, 214102 (2005).

[46] Y. Baskin and L. Meyer, Phys. Rev. **100**, 544 (1955).

[47] Compilation of L. Brewer, Lawrence Berkeley Laboratory Report No. LBL-3720 (unpublished).

[48]  Y. S. Touloukian, R. K. Kirby, R. E. Taylor, T. Y. R. Lee, Thermophysical Properties of Matter, Vol. 13 (Plenum, New York 1977).

[49]  C. Yu, L. Shi, Z. Yao, D. Li, A. Majumdar, Nano Letters **5**, 1842 (2005).

[50]  E. Pop, D. Mann, J. Cao, Q. Wang, K. Goodson and H. Dai, Phys. Rev. Lett. **95**, 155505 (2005).